\documentclass[%
 rsi,
 amsmath,amssymb,
 reprint,%
superscriptaddress,
]{revtex4-1}

\usepackage{graphicx}
\usepackage{dcolumn}
\usepackage{bm}
\usepackage{subcaption}
\usepackage[utf8]{inputenc}
\usepackage[T1]{fontenc}
\usepackage{mathptmx}
\usepackage{xcolor}
\usepackage[font=footnotesize]{caption}
\begin{document}


\title[Sample title]{ Proton deflectometry with \textit{in situ} x-ray reference for 
absolute measurement of electromagnetic fields in high-energy-density plasmas}

\author{C.L. Johnson}
 \email{cj4783@princeton.edu}
 \affiliation{Rowan University, 201 Mullica Hill Rd, Glassboro, NJ 08028}
 \altaffiliation[Presently at: ]{Dept.~of Astrophysical Sciences, Princeton University, Princeton, NJ}

\author{S. Malko}%
 \affiliation{Princeton Plasma Physics Laboratory, 100 Stellarator Road, Princeton, NJ 08540}%

\author{W. Fox}
  \email{wfox@pppl.gov}
 \affiliation{Princeton Plasma Physics Laboratory, 100 Stellarator Road, Princeton, NJ 08540}%
 \affiliation{Dept.~of Astrophysical Sciences, Princeton University, Princeton, NJ 08544}%

\author{D.B. Schaeffer}
 \affiliation{Dept.~of Astrophysical Sciences, Princeton University, Princeton, NJ 08544}%

\author{G. Fiksel}
 \affiliation{Center for Ultrafast Optical Science, University of Michigan, Ann Arbor, MI 48109}%

\author{P.J. Adrian}
\author{G. Sutcliffe}
\author{A. Birkel}
 \affiliation{Plasma Science and Fusion Center, Massachusetts Institute of Technology, Cambridge, MA 02139}%

\date{\today}

\begin{abstract}
We report a technique of proton deflectometry which uses
a grid and an \textit{in situ} reference x-ray grid image for
precise measurements of magnetic fields in high-energy density plasmas. 
A D$^3$He fusion implosion provides 
a bright point-source of both protons and x-rays, which
is split into beamlets by a mesh grid.
The protons undergo deflections as they propagate
through the plasma region of interest, whereas the x-rays travel 
along straight lines.  The x-ray image therefore provides a 
zero-deflection reference image.  The 
line-integrated magnetic fields are inferred from the 
shifts of beamlets between the deflected (proton) and reference (x-ray)
images.
We developed a system for analysis of this data, including 
automatic algorithms to find beamlet locations and 
calculate their deflections from the reference image.
The technique is verified in an experiment performed at OMEGA to measure a non-uniform magnetic field in vacuum, then applied to observe the interaction of an expanding plasma plume with the magnetic field.
\end{abstract}

\maketitle

\section{\label{sec:intro}Introduction}

Magnetic fields are valuable to control and improve energy
confinement in high-energy-density (HED)
plasma experiments.
Strong, 10-100~T scale magnetic fields can be self-generated in the plasma,
or can be applied externally through
pulsed-power technology.
Self-generation processes include the Biermann battery effect~\cite{LiPRL2006,PetrassoPRL2009,YatesPRL1982} and Weibel instability~\cite{FoxPRL2013,HuntingtonNatPhys2015}.
In inertial confinement fusion plasmas, magnetic fields are self-generated in the plasma corona \cite{RyggScience2008} and in hohlraums \cite{GlenzerPoP1999}.
In pulsed-power plasmas, large currents up to 10's of MA (and corresponding magnetic fields) can compress plasma to fusion conditions \cite{SlutzPoP2010,SefkowPoP2014}. 
Both external and internally-generated magnetic fields are important in laboratory astrophysical experiments to investigate 
phenomena including magnetic
reconnection~\cite{NilsonPRL2006,LiPRL2007b,FikselPRL2014,RosenbergPRL2015} and magnetized shocks~\cite{SchaefferPRL2017,SchaefferPRL2019}.

All these applications benefit from advances
in measuring magnetic fields under these plasma conditions,
particularly the development of proton deflectometry (also 
called proton radiography).  
In proton deflectometry, a beam of protons is produced and sent through a plasma; the protons are deflected by the electromagnetic fields within the plasma, and then stream to a detector. 
The electromagnetic
fields are inferred from analyzing the final proton positions
on the  detector.
In many experiments~\cite{NilsonPRL2006,LiPRL2006,LiPRL2007b,PetrassoPRL2009,WillingalePRL2010,RosenbergNatComm2015}, a mesh is used 
to split the proton
beam into a number of beamlets to more easily identify deflections.  The mesh-based proton deflectometry 
has been applied to measure the dynamics of Biermann battery fields
\cite{PetrassoPRL2009, WillingalePRL2010},
as well as the interaction and magnetic reconnection between
colliding Biermann fields \cite{LiPRL2007b, RosenbergNatComm2015}. 

A limitation in applying this basic technique to
new experiments is that it requires a region of zero magnetic field
in the deflection image in order to determine
the zero-field reference pattern
of the mesh.   This limits the
use when the magnetic fields fill the measurement volume, 
particularly for the case with a global magnetized volume, such
as applied by pulsed-power technology.
This calls for an expansion of the measurement technique.
Accordingly, 
in this paper, we develop and describe a novel proton deflectometry technique
that is suitable for measuring electromagnetic
fields in globally-magnetized plasmas.
The main advancement is the addition of an
image plate x-ray detector to the detector stack;
since x-rays are not affected by electromagnetic fields, the shadow recorded on the image plate can provide the reference image of the grid suitable
for direct calculation of proton deflections.
This method is also superior to obtaining
reference grid images from surrogate shots 
(without plasma and magnetic field),
as it does not depend on the reproducibility of
the mesh alignment between shots.
We describe the implementation of this measurement technique
at the OMEGA laser facility using a D$^3$He fusion backlighter~\cite{LiPRL2006,PetrassoPRL2009} to measure magnetic fields in strongly-magnetized
static (no plasma) and dynamic (with plasma) experiments.

The paper is organized as follows: Section II describes the basic principles of proton deflectometry to motivate the required measurements; Section III describes the setup and application of the technique at OMEGA; 
Section  IV explains the step-by-step method for analysis of the obtained data including image processing, contrast enhancing, and obtaining the line-integrated magnetic field; Section V shows an example of the experimental data obtained using this technique to study magnetic field dynamics; finally, we make conclusions in Section VI.


\section{\label{sec:prad}Proton deflectometry measurement}

\begin{figure}
\includegraphics[width=\columnwidth]{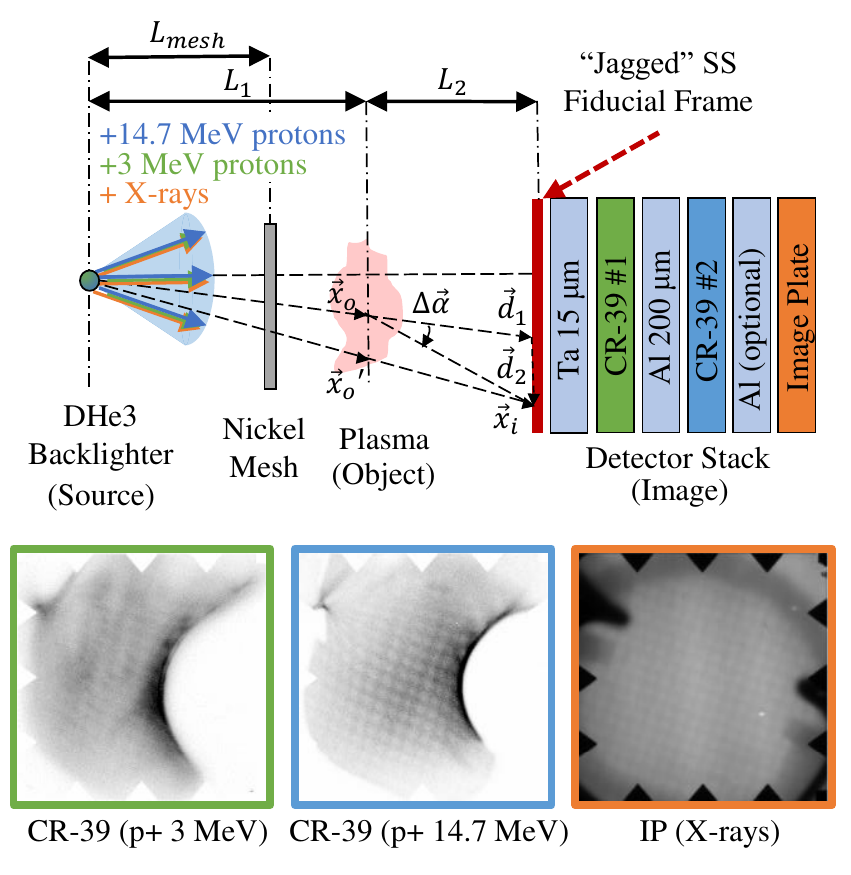}
\caption{\label{fig:prad_setup}The typical proton radiography setup consists of a proton source which produces a stream of protons and x-rays that is split into beamlets by a mesh. The protons are deflected by the electromagnetic fields in the plasma and the deflected beamlet positions are recorded on a detector. The detector stack used in the experiment at OMEGA consists of two CR-39 detectors to record 3 MeV and 14.7 MeV proton positions and an image plate to record the undeflected x-ray shadow of the mesh. Example data from the two CR-39 detectors and the IP are shown. (Not to scale)}
\end{figure}

In this section we recapitulate the proton deflectometry theory \cite{KuglandRSI2012}
to motivate the experimental setup and required measurements.
The basic proton radiography setup is shown in 
Fig.~\ref{fig:prad_setup}.  
Here we primarily consider a fusion implosion D$^3$He backlighter \cite{SeguinRSI2003}, which produces
a bright point source of protons with birth energies at 3 and 14.7 MeV
from the D-D and D-$^3$He fusion reactions.
The implosion is also a bright point-source of x-rays with energies on the order of 10's of keV.
The protons and x-rays are split into discrete beamlets using a mesh.
When the protons travel through the magnetized plasma volume, they are deflected
by the electromagnetic fields in the plasma, 
and the final proton tracks are recorded on a detector.  Of course, the x-rays are undeflected
and therefore maintain the initial beamlet pattern to the detector.
The system is fielded in a magnified point-source geometry,
where the magnification from the ``plasma'' or object plane to
the detector plane is $M = (L_1 + L_2) / L_1$.
The detector stack shown records the 3 and 14.7~MeV protons
and x-rays, and is discussed in greater detail below.

As the protons traverse the plasma, they are deflected according to the Lorentz force law,
\begin{equation}
    \frac{d\vec{v}}{dt} = \frac{e}{m_p} (\vec{E} + \vec{v} \times \vec{B}).
    \label{EqLorentz}
\end{equation}
Consideration of the limit of small deflections (owing to the high proton energy)
and the  paraxial approximation discussed in Ref.~\cite{KuglandRSI2012} where the distance $L_1$ is much larger than the object (plasma) size, one can approximate the solution to Eq.~\ref{EqLorentz}
as an integral over the straight-line proton trajectory through the plasma.
The proton receives a simple impulsive angular deflection $\Delta \vec{\alpha}$ as it passes through the plasma, after which it propagates along a straight-line to the detector.  The deflection is given in this limit by an integral along the 
proton path,
\begin{equation}\label{eq:alpha}
\Delta \vec{\alpha}_{E,B}=\frac{e}{m_p v_p^2}\int{(\vec{E}+\vec{v}\times\vec{B}) dl},
\end{equation}
where $v_p = \sqrt{2 E_p / m_p}$ is the proton velocity given
its energy $E_p$.
This deflection angle includes contributions from both electric and magnetic fields. Often, as in the present case, the electric field contribution
can be ignored and the deflection is given by
\begin{equation}\label{eq:alpha_B}
\Delta \vec{\alpha}_{B}=\frac{e}{m_p v_p}\int{d\vec{l} \times \vec{B} }.
\end{equation}

This result is valuable as it directly relates the angular deflection of 
the protons to the line-integrated magnetic field, which is the quantity of
interest for measurement. 
Since the magnetic fields vary throughout the 
plasma, the deflection angle $\Delta \vec{\alpha}$ is a
function of the position $\vec{x_o}$ at which the 
proton beamlet crosses the plasma.  Accordingly, by obtaining the
deflection angle as a function of position using many beamlets, one can construct a 2-D map
of the line-integrated field vs. position. 

Considering a beamlet which intersects the plasma at position $\vec{x_o}$,
it will be deflected by the angle $\Delta \vec{\alpha}_B$ and propagate to the detector, where its
final position in the image plane $\vec{x_i}$ is 
\begin{equation}\label{eq:defl_pos}
\vec{x_i}= \vec{d_1}+\vec{d_2}=M\vec{x_o}+L_2\Delta\vec{\alpha_{B}},
\end{equation}
where $\vec{d_1}$ is the undeflected proton position in the detector plane accounting 
simply for the magnification $M$,
and $\vec{d_2}$ results from the deflection in the plasma.  
Therefore, if the undeflected $M\vec{x_o}$ and deflected $\vec{x_i}$ positions are measured in the detector plane, the line-integrated magnetic field can be calculated directly using
\begin{equation}\label{eq:implane_Bdl}
\int{d\vec{l}\times\vec{B}} = \frac{m_p v_p}{e} \frac{1}{L_2} (\vec{x_i} - M\vec{x_o}).
\end{equation}

However, since $\int d\vec{l} \times \vec{B}$ is a function of the plasma plane position $\vec{x_o}$,
it is often most convenient to work in strictly plasma-plane coordinates. 
Therefore, Eq. (\ref{eq:implane_Bdl}) can be rearranged by introducing 
 $\vec{x_o}'=\vec{x_i}/M$, which is the deflected position, registered back 
to the plasma plane.  Then, 
\begin{equation}\label{eq:final_Bdl}
\left(\int{d\vec{l}\times\vec{B}}\right) _{\vec{x_o}} = \frac{m_p v_p}{e} \frac{L_1+L_2}{L_1 L_2} (\vec{x_o}' - \vec{x_o}).
\end{equation}

This equation outlines the required measurements.  For each beamlet, we require a measurement of its final position $\vec{x_o}'$ as well as its undisturbed position $\vec{x_o}$.
Of course, 
the proton detector directly indicates the final positions $\{\vec{x_o}'\}$ for each beamlet,
but it does not directly provide $\{\vec{x_o}\}$
We note that previous experiments have inferred the $\{\vec{x_o}\}$ for the beamlets 
by extrapolating from mesh areas where the magnetic field is assumed zero or small,
and fitting an undisturbed mesh throughout the rest of the domain \cite{LiPRL2006, WillingalePRL2010}.  However, this method cannot be applied to globally magnetized experiments.  In these cases, $\vec{x_o}$ could
potentially be found by using a $B=0$ calibration shot, but this depends on consistent mesh 
construction and alignment between shots.  This motivates the 
present method where we use the x-rays from the proton source to directly produce a shadow of the mesh and indicate $\vec{x_o}$ for each beamlet.
Note, a final requirement is that, given multiple beamlets on the detector, we also must unambiguously match the deflected and undeflected pairs.  This requires adding some simple fiducial elements to the grid as is discussed below.

Finally, we calculate some
useful quantities
related to spatial resolution and measurement limits.
These quantities all depend on the magnification
setup of the mesh, plasma,
and detector, and can be
tailored in future experiments as needed.
The spatial resolution of the beamlet measurement
relative to the plasma is set by the mesh period
in the plasma plane: $\Delta x_{mesh,o} = M_{mesh} \Delta x_{mesh} $, 
where $M_{mesh}$ is the magnification of
the mesh to the object plane $(L_1 / L_{mesh})$, 
$\Delta x_{mesh}$ is the physical mesh period,
and $L_{mesh}$ is the distance from the backlighter to the mesh.
Next, as discussed in detail below, we find that the present
accuracy of the measurements is approximately at the one-pixel
scale at the detector.  The equivalent
$\delta Bdl$ for one pixel deflection $\delta x_{1,d}$ is 
$(\delta x_{1,d}/L_2) (m_p V_p/e)$
and is therefore a representative measurement uncertainty.
Finally, we consider the deflection associated with
shifting the beamlets an entire mesh period: $\delta B dl$ = 
$(\Delta x_{mesh} / L_{mesh}) (L_1+L_2)/L_2 (m V_p /e)$.
This is a useful quantity as it indicates the scale where
the magnetic deflections are becoming ``large'' and
beginning to lead to significant mesh deflection
and distortion.  As found below, 
we are able to measure deflections somewhat larger than a one-beamlet
shift; nevertheless, it is recommended that measurement setups should be chosen 
so that $B dl$ not exceed this by
more than a factor of a few, since the beamlet association
between deflected and undeflected positions will quickly become ambiguous.

\begin{figure}
\includegraphics[width=0.8\columnwidth]{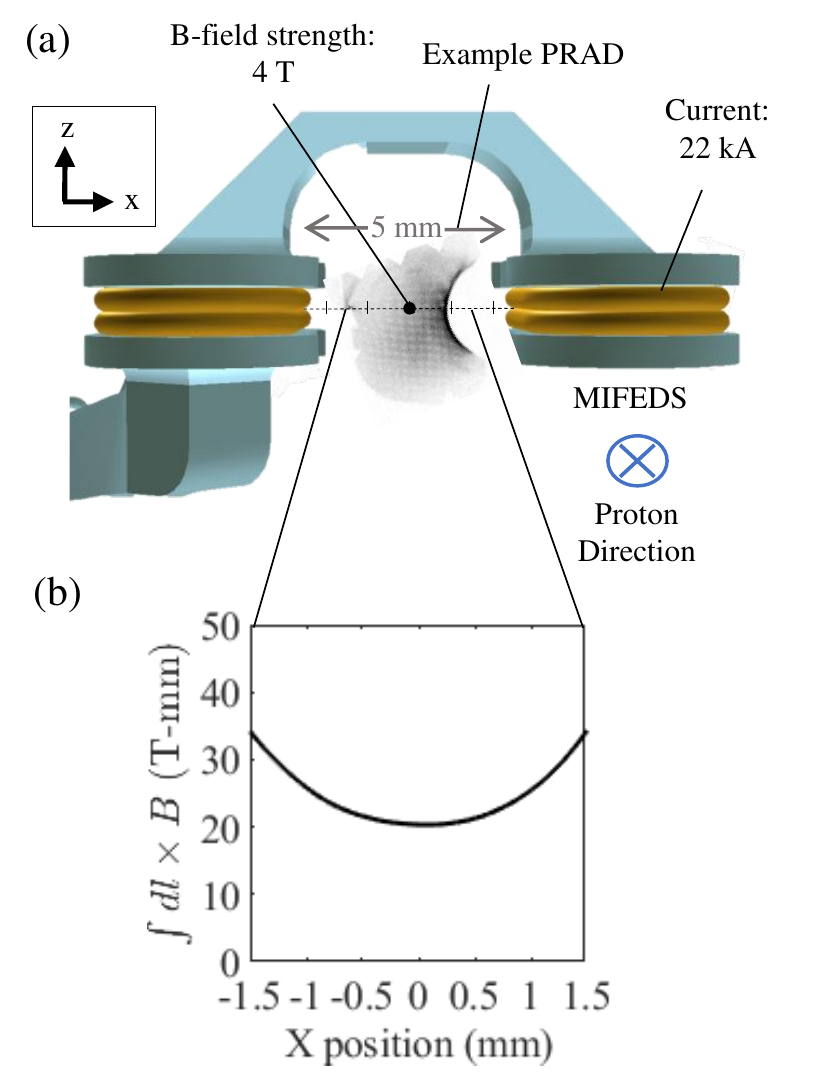}
\caption{\label{fig:vac_setup}
(a) Schematic of experimental setup (not to scale) for vacuum magnetic field validation measurement at OMEGA with example PRAD. (b) COMSOL-calculated vacuum path-integrated magnetic field profile for the area between the MIFEDS coils probed by the proton beam ($\pm1.5$ mm from center). 
}
\end{figure}

\section{\label{sec:experiment} Implementation at OMEGA}

This proton radiography technique was applied to experiments performed at the OMEGA 60 laser facility \cite{BoehlyRSI1995}, using imploded D$^3$He-filled capsules to
produce point-source 3 MeV and 14.7 MeV protons as fusion products,
as well as a bright source of x-rays \cite{SeguinRSI2003}.  As motivated above,
we augment a standard detector stack with an additional image plate
to detect a x-ray shadow of the mesh, to provide absolute reference positions of the 
mesh.  This section describes in detail the setup of the backlighter, mesh, and detector stack, and then measurements of the magnetic fields in a quasi-static (no plasma) demonstration experiment.

The capsules were 420 $\mu$m in diameter and were filled with a
mixture of D$_2$ and $^3$He at equimolar pressures of 14.1 and 6.6 atm, respectively.
The capsule was imploded by 19 OMEGA drive beams, each delivering an energy of 500 J
in a 1 ns square pulse.  The lasers did not use a phase plate, and were
slightly defocused from the capsule, with the focal point set to 
a distance of $\sim9 R = 1.82~\mu$m short
of the capsule, where $R$ is the capsule radius.
Nuclear measurements on the experiments indicated neutron yields of $0.5$-$2\times10^{9}$ into
$4\pi$, with burn-averaged ion temperatures of
7.7-9.5~keV.
The D$^3$He $14.7$ MeV proton yields 
over the experimental day ranged from $3.3\times10^{8}$-$1.4\times10^{9}$ protons into $4\pi$, and in particular the yield was $1.2\times10^9$ 
for the shot analyzed in Section~\ref{sec:analysis},
which yielded an average of 104 protons/pixel on the detector.
The capsule implosion also produces a bright point source of x-rays with characteristic energies of 10's of keV, comparable to the ion temperature.
The backlighter capsule was positioned 
at an offset $L_1 =$~10~mm from the target-chamber-center (TCC), where the TCC
defines the object plane, and the detector stack was positioned a distance
$L_2 =$~154~mm opposite.

A Ni mesh with 125 $\mu$m pitch, 90 $\mu$m mesh opening and bar thickness of 35 $\mu$m was used to split the proton beam into beamlets.
The mesh was positioned 4 mm from the backlighter source, so that the
mesh period in the object plane was 312 $\mu$m.  The mesh causes an energy
downshift and scattering of protons, and attenuation of x-rays, which leaves
an imprint in the proton and x-ray image data.

The detector stack (Fig.~\ref{fig:prad_setup}) consisted of two CR-39 detectors for
the 3~MeV and 14.7~MeV protons, and an added image plate detector (IP, Fujifilm SR-type)
primarily sensitive to x-rays, which records an x-ray shadow of the mesh.  
The CR-39 detectors were processed and scanned under a microscope 
by established techniques \cite{SeguinRSI2003}, producing a
map of proton counts per pixel.
The IP was scanned with commercial Fujifilm Typhoon FLA-7000 scanner 
with resolution of r = 25 $\mu$m and sensitivity of $S$ = 1000.
The Ta, Al, and CR-39 pieces
filter the x-rays reaching the IP.  
For the given stack with 15~$\mu$m Ta, 200~$\mu$m Al, and $2\times1500$~$\mu$m of CR-39, the minimum energy reaching the IP is $h\nu \gtrsim~25$~keV (at $1/e$). We obtained the best contrast performance at OMEGA without the second (optional) Al filter, though it may be used to limit signals onto the IP if needed.
We also introduced a jagged fiducial frame in the front of the stack which
leaves a "tooth" pattern around the border that facilitates the alignment
of the CR-39 and IP images in later data processing.


 We first conduct a validation experiment to demonstrate this technique to measure a 2-D map of 
 a static magnetic field. The experiment used the MIFEDS (Magneto-Inertial Fusion Electrical Discharge System) \cite{FikselRSI2015} pulsed power system.
 The coil design supplies an open experimental geometry which is useful
 for experiments (and was previously used in Ref.~\cite{FikselPRL2014}), 
 and is somewhat non-uniform as the field strength increases
 toward the coils on the left and right sides of the measurement volume.
 Fig.~\ref{fig:vac_setup}a shows the experimental setup, with an accompanying
 oriented PRAD image. 
 Examples of the full set of vacuum magnetic field shot CR-39 and IP data are shown in Fig.~\ref{fig:prad_setup}. 
 The typical magnetic field strength at the midpoint between the coils was 4 T
over the characteristic volume of (5 mm)$^3$. The corresponding MIFEDS current of 22 kA was used for the calibration shots. The coil was modeled using 
COMSOL to obtain the path-integrated magnetic field relevant for
proton deflections (Eq.~\ref{eq:alpha_B}) for the region between the coils (Fig.~\ref{fig:vac_setup}b).
Relevant values for the experimental geometry are summarized in Table~\ref{table:geometry}.



\begin{table}
\begin{center}
\begin{tabular}{lc}
\hline
Source-object separation $L_1$                  & 10 mm \\
Object-detector separation $L_2$                & 154 mm \\
Magnification  $M$                                 & 16.4 \\
Field of view at object plane                   & 6.1 mm \\
Pixel size at detector plane $\delta x_{1,d}$                      & 314.5 $\mu$m \\
Pixel size in object plane $\delta x_{1,o} = \delta x_{1,d}/M$        & 19.2 $\mu$m \\
Mesh period  $\Delta x_{mesh}$                                   & 125 $\mu$m \\
Mesh period at object plane $M_{mesh}\Delta x_{mesh}$        & 312 $\mu$m\\
Equivalent $\delta Bdl$ of one pixel deflection & 0.9 T mm \\
Equivalent $\delta Bdl$ of one mesh unit        & 18.5 T~mm\\
\hline
\end{tabular}
\caption{Relevant experimental geometry values.}
\label{table:geometry}
\end{center}
\end{table}

\section{\label{sec:analysis} Proton Deflection Analysis}

\subsection{\label{sec:image}Image Processing}

\begin{figure}
\includegraphics[width=\columnwidth]{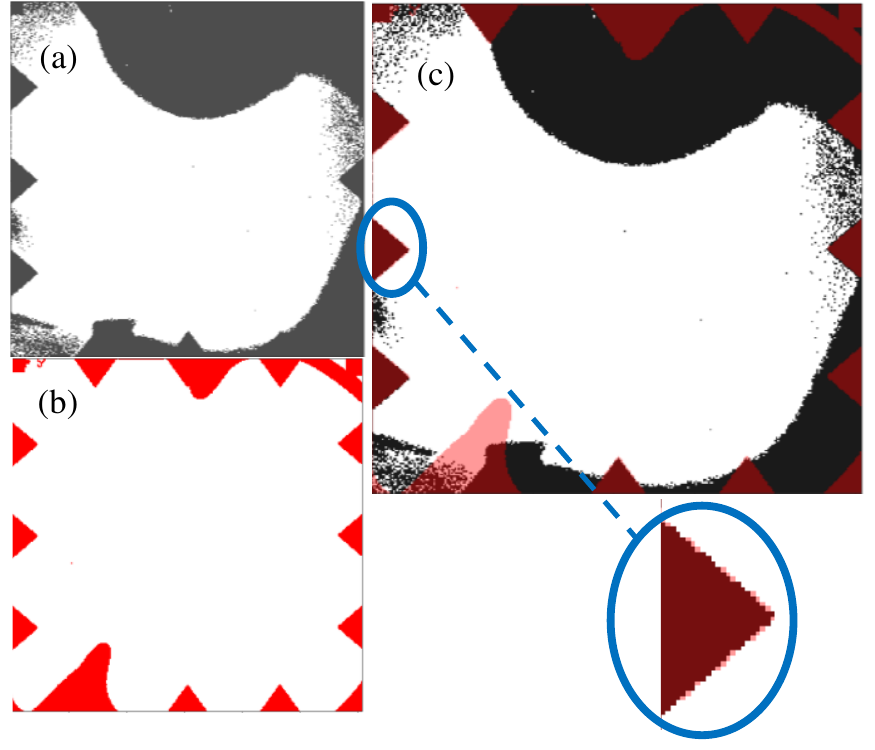}
\caption{\label{fig:align} Image alignment based
on teeth features imprinted by the fiducial frame.  
The CR-39 (a) and IP (b) images are masked to isolate the teeth.
The masked IP image is rotated and scaled to exactly match the CR-39 image (c).}
\end{figure}

\subsubsection{\label{sec:align}Aligning IP and CR-39}
In order to compare the IP and CR-39 to calculate the beamlet deflections, the images must be the same size and aligned to produce a 1:1 pixel comparison.
The IP image has a higher resolution than the CR-39 and therefore must be cropped to isolate the image, eliminating extraneous boundary pixels beyond the frame, and resized to match the size of the corresponding CR-39. For the shots in which a jagged fiducial frame was used, the teeth from the IP and CR-39 images were isolated and the images overlapped to facilitate alignment (Fig.~\ref{fig:align}). The IP often required corrections including cropping, flipping, and/or rotating in order to achieve alignment with the corresponding CR-39. For shots where no jagged frame was present, we aligned the IP to the CR-39 based on the identified frame features, though
this was more laborious and error prone due to the difficulty in identifying the features in all images, resulting in error of $\pm$ 1 pixel (therefore $\pm$ 0.9 T~mm). 

\begin{figure}
\includegraphics[width=\columnwidth]{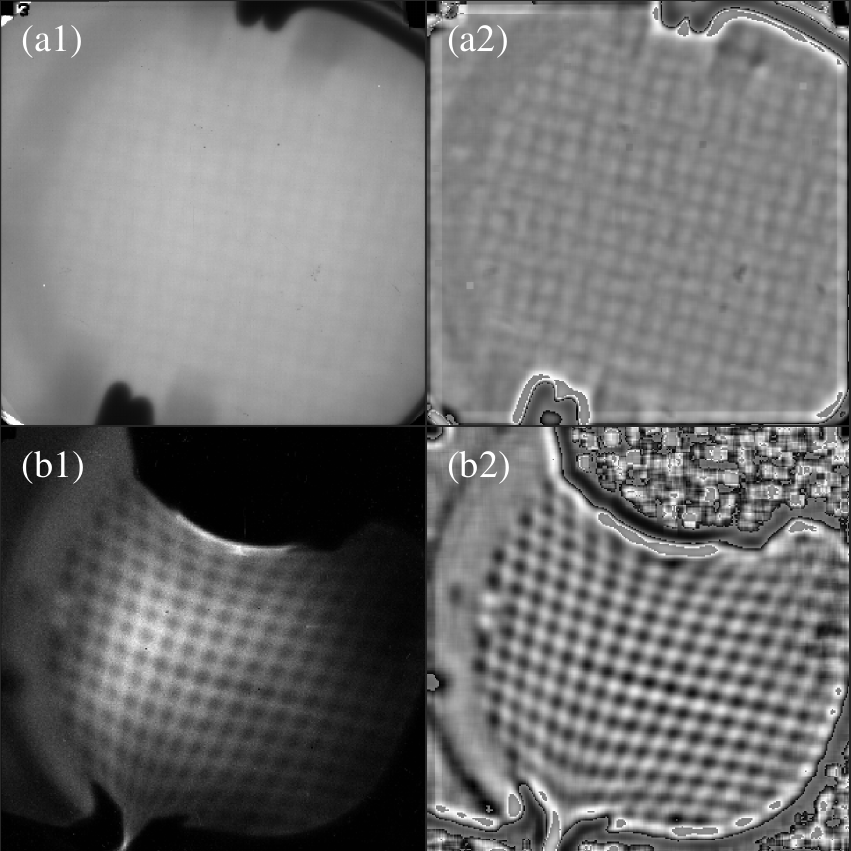}
\caption{\label{fig:enhance} Comparison of original (1) and enhanced (2) IP (a) and CR-39 (b) images. Filters are applied for smoothing and normalizing proton fluence across the image, then the contrast is improved in the mesh area of the image.}
\end{figure}

\subsubsection{\label{sec:enhance}Enhancing Images}
To ease the process of finding beamlet locations, the images were enhanced
to reduce the noise and improve the contrast (Fig.~\ref{fig:enhance}). 
First, a ``box-car'' filter with a size of 10x10 pixels (approximately the size of one beamlet) was applied to smooth the images. Then, another box-car filter of size 20x20 pixels was applied, and divided out from the smoothed image to normalize the fluence. The maximum and minimum pixel values within the mesh area of the image were used to define the final range of pixel values in the image; any pixel values in the image outside of this range were re-scaled.  This ensures that the mesh is the area of the image with the maximum contrast.

As a figure of merit, we evaluated a ``beamlet contrast'' in the images, which we defined as the ratio of the maximum/mininum range of the beamlets
to the rms of the sub-beamlet-scale noise (which can disrupt identifying the beamlet location).  We found that this enhancement process results in a contrast improvement on the vacuum CR-39 images from $\sim$8 on the original to $\sim$60 on the enhanced.

An additional filter was applied to the IP image in order to 
further reduce the noise within each mesh opening, 
making the center of the mesh opening the location of the local extremum to aid in later processing of beamlet locations. 
The filter mimicks the shape of a beamlet, i.e. low values around the 2-pixel edges and high values in the 5x5 pixel center. This process results in an improvement of contrast for the IP from $\sim$2 on the original to $\sim$19 on the enhanced.

\subsection{\label{sec:beamlet}Beamlet Detection}
Once the contrast is improved, the images are rotated so that the rows are approximately horizontal, which allows an
automated code to move along the row and search for the local extrema that correspond to beamlet centers. For ease of discussion, a second coordinate system in pixel units is defined in the frame of the image where $u$ is along the row and $v$ is along the column, i.e. in the rotated image $u$ is along the horizontal and $v$ is along the vertical.

This code takes user inputs that set the parameters for each row to be analyzed. These parameters include the indices of the beginning and end of the row in $u$, the indices of the bottom and top of the row in $v$, a beamlet spacing parameter, and row/column identifiers.

\begin{figure}
\includegraphics[width=\columnwidth]{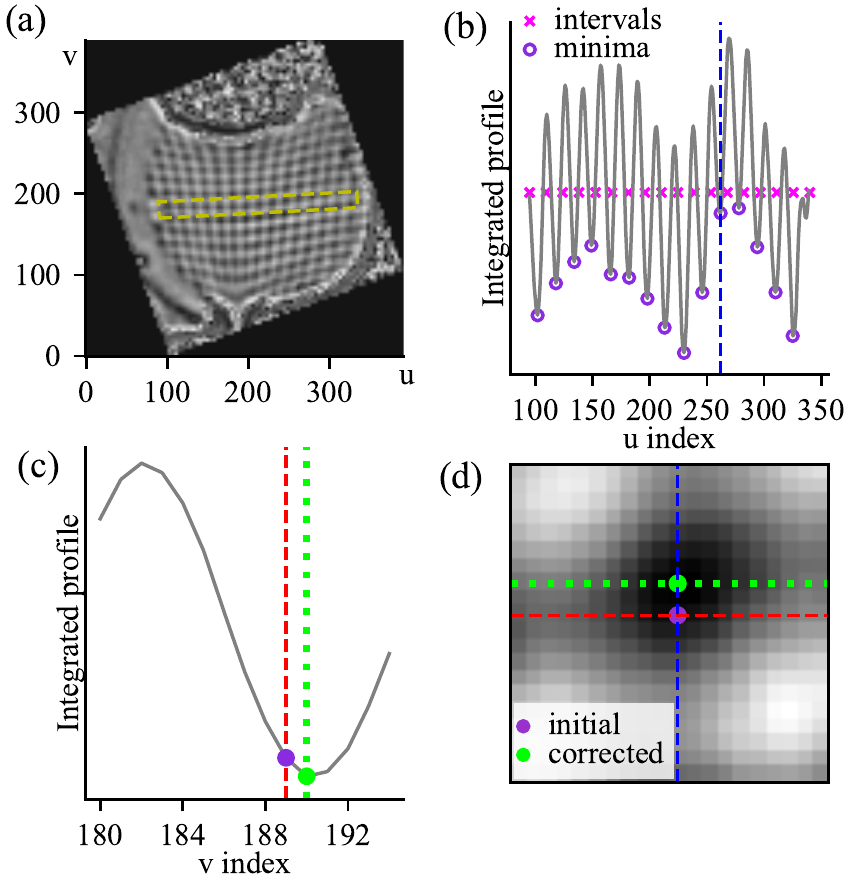}
\caption{\label{fig:aut_code} Detection of beamlet coordinates (a) The input parameters define the area occupied by the desired row. (b) A line-out along $u$ of pixel values integrated along $v$ is parsed in intervals defined by the beamlet spacing input parameter. The minimum (CR-39) or maximum (IP) in each interval is found and recorded as the initial $u$-coordinate. 
(c) For each $u$-coordinate, the $u$-integrated profile is determined for the $v$-values in the defined area. The appropriate extremum of this profile is recorded as the initial $v$-coordinate. (d) The $u$- and $v$-coordinates define a starting point that is corrected after the checks are performed so that the actual local extremum is identified for each beamlet.}
\end{figure}

\begin{figure}
\includegraphics[width=0.7\columnwidth]{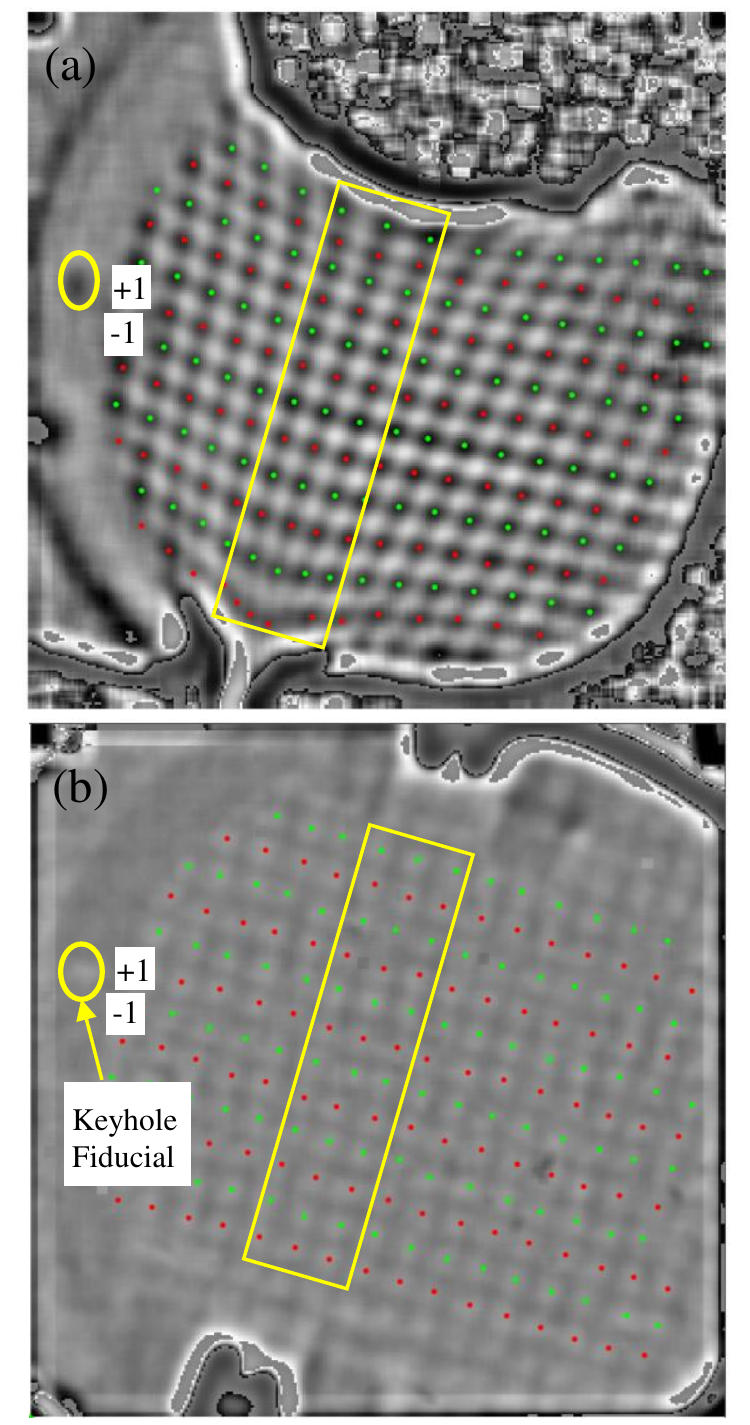}
\caption{\label{fig:final_im} Final processed CR-39 (a) and IP (b) images with all beamlets identified. The keyhole fiducial is used for identifying the rows and columns in order to compare corresponding beamlets between images. The three columns boxed lie between the MIFEDS and therefore define the area of interest for calculating the line-integrated magnetic field. }
\end{figure}

The $v$ start and end index values are averaged to set the initial $v$-index for which the line-out along the row is drawn (Fig.~\ref{fig:aut_code}a). This line-out is used to find the local extrema and therefore the initial index values of the beamlet coordinates in $u$  (Fig.~\ref{fig:aut_code}b).  For each $u$-value, the pixels above and below the averaged $v$-value are searched to find the extremum in the $v$-direction (Fig.~\ref{fig:aut_code}c). These $u$ and $v$ values set a preliminary coordinate for the beamlet that is then corrected and adjusted by searching around the initial point to find the actual local extremum  (Fig.~\ref{fig:aut_code}d).

The code also includes built-in checks that combat some common errors in the automatic location of the beamlets. These include checking the spacing of the beamlets in $u$ to prevent double-counting of beamlets, checking the distances between consecutive $v$ values to prevent jumping to a different row as can result in cases of large deflections, and repeatedly checking the pixels around each coordinate to ensure that the actual local extremum has been identified.

The final corrected coordinates for the analyzed row are saved into a data structure that records points by row and column index. These indices are defined using the fiducials on the nickel mesh including a keyhole cutout and glue dots over the mesh. These mark where row indices +1 and -1 and column index 0 are defined as shown in 
Fig.~\ref{fig:final_im}. The data structure is exported to a file.  
Once all coordinates have been found, they are transformed back to the initial unrotated, aligned images.
  
Despite the built-in checks, occasionally some points are not centered in the beamlets,
so it remains important to spot check the beamlet locations against the raw data and 
apply manual adjustments when needed. 
In general, this entire process results in an error of $\pm$ 1 pixel for the beamlet locations.  A completed IP and CR-39 after both the automatic code and manual adjustments is shown in Fig. \ref{fig:final_im}.

\subsection{\label{sec:calculation}Line-Integrated Magnetic Field}
\begin{figure}
\includegraphics[width=0.95\columnwidth]{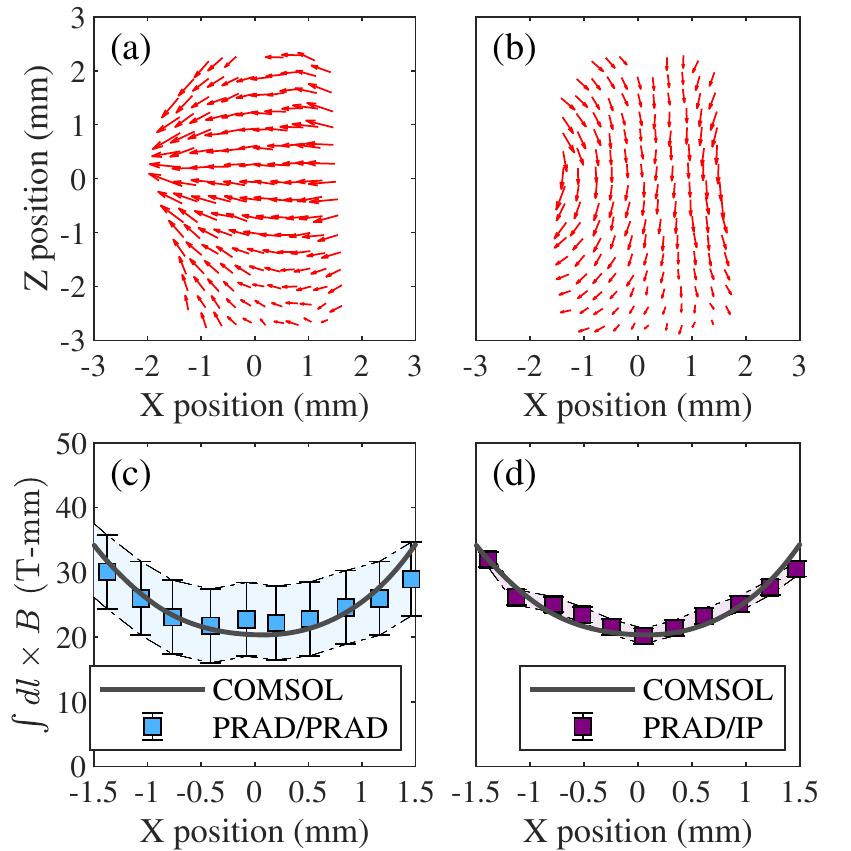}
\caption{\label{fig:Bdl} Process of calculating line-integrated magnetic field. (a) Spatial distribution of beamlet deflections, (b) Spatial distribution of $\int{d\vec{l}\times\vec{B}}$, (c) Final profile of line-integrated magnetic field versus position between the MIFEDS for PRAD/PRAD technique and (d) PRAD/IP technique.}
\end{figure}

Once corresponding IPs and CR-39s have been analyzed and the coordinates of the beamlets have been found, the path-integrated magnetic field can be calculated from the deflections. This process was completed for a calibration shot for the magnetic field in vacuum using the 14.7 MeV proton radiography image.  We directly compare the CR-39 deflected beamlet coordinates with the undeflected beamlet coordinates on the IP; the spatial distribution of deflections is plotted in Fig.~\ref{fig:Bdl}a. The deflections are then used to calculate the corresponding line-integrated magnetic field at each beamlet location using Eq. (\ref{eq:final_Bdl}).  The line-integrated magnetic field map is plotted in Fig.~\ref{fig:Bdl}b and shows that in this highly-magnetized system, there is no area of zero magnetic field that can be used to provide a reference for analyzing the rest of the proton radiography image. This is where using the IP is necessary to provide reference positions. Finally, the magnitude of the path-integrated magnetic field for the three columns in the midplane between the MIFEDS (as outlined in Fig.~\ref{fig:final_im}) can be averaged and plotted as a function of position (Fig.~\ref{fig:Bdl}d). The experimental data is also compared to the COMSOL Biot-Savart calculations of the magnetic field based on the coil geometry.

In order to evaluate the precision and efficiency of the new proposed technique using the IP, we compared the results to those obtained by an alternative 
method in which the reference undeflected beamlet positions were obtained from a CR-39 image of a separate vacuum shot with no magnetic field applied. The line-integrated magnetic field calculated using this technique is shown in Fig.~\ref{fig:Bdl}c. Comparing both results, one can see that the PRAD/IP technique improves the accuracy of the magnetic field measurement by a factor of $\sim$4. 
The resulting error bar for both cases consists of the standard deviation of the path-integrated magnetic field values for the three central columns $\sigma$ and the systematic error $\sigma_{sys}$ from the alignment of the data images. The alignment error for the PRAD/PRAD technique has two sources: the error from overlapping the two CR-39's, $\pm$ 2 pixels, and the alignment error of mesh positioning in the experimental setup between the two separate shots that is estimated to be 25 $\mu$m and yields $\sim\pm$ 3 pixels in the detector plane.
The PRAD/IP alignment error is $\pm$ 1 pixel due to the high resolution of the IP image. Then the total error is calculated as $\sigma_{tot}$ = $\sqrt{\sigma^2 + \sigma_{sys}^2}$. The PRAD/IP technique yields an average error of $\sim$1.2 T~mm, while PRAD/PRAD yields an error of $\sim$5.7 T~mm. The error bar for the new technique can be further improved by increasing the magnification, using the jagged frame to reduce alignment error, using other mesh periods, and using a mesh material of higher Z in order to enhance the contrast and therefore improve the beamlet identification in the automatic routine. 

\begin{figure}
\includegraphics[width=\columnwidth]{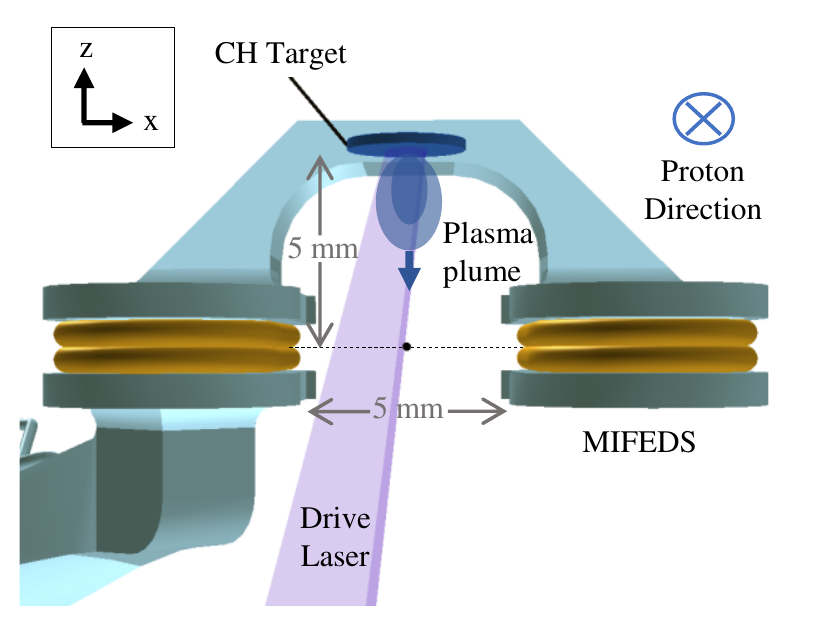}
\caption{\label{fig:exp_setup}The setup of the experiment performed at OMEGA used MIFEDS coils to  produce a magnetic field in the $-z$-direction. The CH-target was ablated to produce a plasma plume also expanding in the $-z$-direction. The proton radiography setup is along the $y$-direction and produces images of the $x$-$z$-plane.}
\end{figure}

\begin{figure*}
\includegraphics{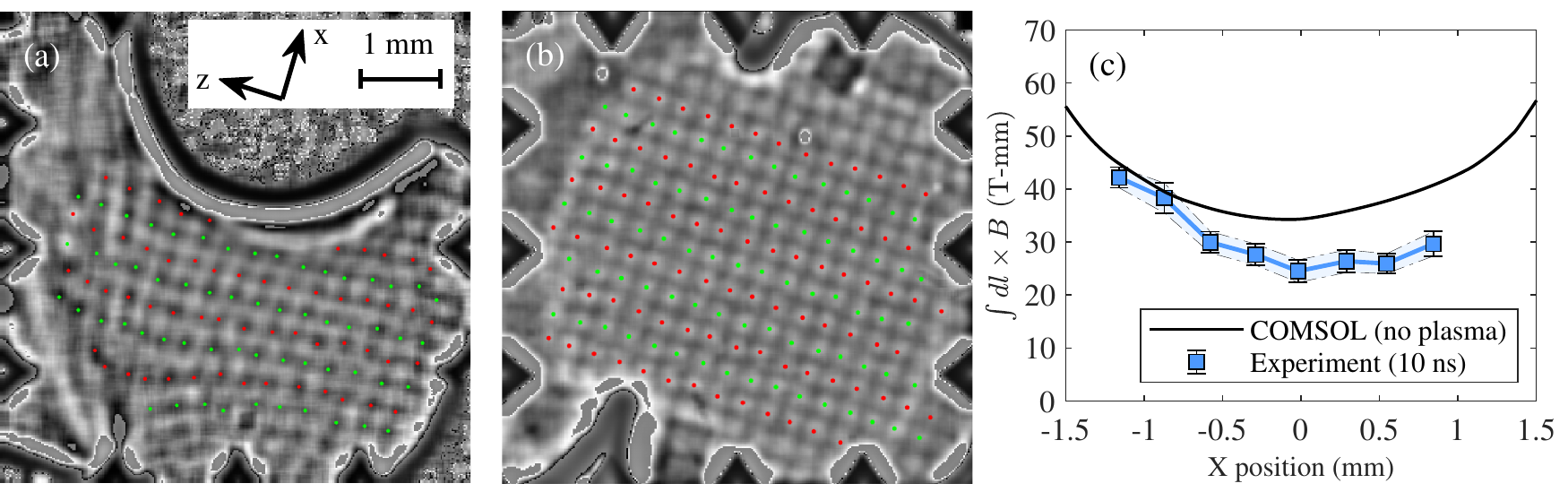}
\caption{\label{fig:Plasma_Bdl} Example of processed data from a plasma shot at 10 ns. (a) CR-39 with beamlet locations identified. 1 mm scale indicates units in the plasma plane. (b) Corresponding IP with beamlet locations identified. (c) Line-integrated magnetic field at this time point compared to the vacuum magnetic field provided by the COMSOL calculation; this shows that by 10 ns some of the magnetic field had been expelled from between the MIFEDS by the plasma plume }
\end{figure*}

\section{Application to expanding plasma experiment}

This technique was employed in an experiment at the OMEGA laser facility to study magnetic field dynamics in HED plasma in the $\beta>1$ regime including the magnetic field evolution over time, its diffusion, and potential anomalous transport processes. 
Expansion of the plasma into the background magnetic field
locally expels the magnetic field and produces a diamagnetic cavity \cite{DimontePRL1991,RipinPOFB1993,CollettePOP2011,NiemannPOP2013,WinskeFASS2019}.
The experimental setup is shown in Fig.~\ref{fig:exp_setup}. 
The coil geometry was identical to the calibration
experiments discussed above; however, the MIFEDS current was increased to 34 kA.  A plastic (CH) target was ablated to produce a plasma plume that flowed parallel to the field and into the region between the coils.
The source target was 5~mm from the center of the diagnosed region.
The laser energy on-target was 20 J in a 1 ns square pulse, the wavelength was 351 nm, and
the laser-incidence angle on target was 33$^{\circ}$.
The plasma was additionally diagnosed using $2\omega$ Thomson scattering to measure plasma parameters, such as electron temperature and density.

 As plasma expands into the magnetic field, it tends to push the magnetic field out of the volume it occupies leading to the formation of a diamagnetic cavity. 
This experiment was particularly focused on the study of diamagnetic cavity formation and evolution in time, which requires a precise magnetic field measurement. 
In Fig.~\ref{fig:Plasma_Bdl} we show an example of the proton radiography data with 14.7 MeV protons (CR-39) and the corresponding x-ray image of the mesh (IP) obtained at 10 ns plasma expansion time. Fig.~\ref{fig:Plasma_Bdl}c shows the resulting line-integrated magnetic field evaluated along the midplane between the coils. One can clearly see the beginning of magnetic field cavitation by comparing the experimental data to the COMSOL calculation of the background magnetic field with no plasma. The CR-39 for this plasma shot shows an instance where a caustic 
(a locus of proton focusing and therefore beamlet overlap) interferes with the data \cite{KuglandRSI2012}; the three rows closest to one of the coils are cut off by the caustic, limiting the amount of data available for analysis.

\section{\label{sec:discussion}Discussion and Conclusions}

In conclusion, a novel technique of proton radiography was developed
which simultaneously obtains a reference image of the undeflected beamlet
locations by using an image plate x-ray detector in the detector stack. 
This technique allows absolute measurement of magnetic fields
in systems which are globally magnetized and have no regions
of zero magnetic field.  Compared to a technique measuring
the mesh pattern on surrogate experiments with $B=0$ and no plasma,
the measurement is more accurate, largely owing to eliminating
the uncertainty regarding reproducibility of the mesh
construction and alignment between shots, 
and we estimate a factor $\sim$4 improvement for the present parameters.
For efficient CR-39 and IP image processing we developed an automatic routine that allows for quick and accurate positioning of the beamlets.
Using this method, we also have performed a characterization of the diamagnetic cavity formation and evolution in time in a $\beta$>1 expanding plasma.

The data that support the findings of this study are available from the authors upon request.

\section{Acknowledgements}

This work was supported through DOE Laboratory Directed Research and Development.  PJA, GS and AB were supported by the U.S. Department of Energy under Grant No. DE-NA0003868. PJA was also supported under Grant No. DE-NA0003960.  The experiment was conducted at the Omega Laser Facility with the beam time through the Laboratory Basic Science under the auspices of the U.S. DOE/NNSA by the University of Rochester’s Laboratory for Laser Energetics under Contract DE-NA0003856.

\bibliography{refs}

\end{document}